\documentstyle[12pt]{article}
\begin{document}
\title{
Predictive Ansatz for Fermion Masses in SUSY GUTS}
\author{
G. K. Leontaris\\
Theoretical Physics Division\\
University of Ioannina\\
GR-451 10 Ioannina\\
Greece
\and
N. D. Tracas\\
Physics Department\\
National Technical University\\
GR-157 80 Zografou\\
Athens, Greece}
\date{}
\maketitle
\begin{abstract}
\noindent
We reexamine a succesful fermion mass Ansatz proposed by Giudice for a wide
range of the ratio $tan\beta =\frac {<\bar h>}{<h>}$ (where ${\bar h},h$ are
the two standard higgs fields), in the context of supersymmetric
grand unified theories. We find that the 7 predictions of the ansatz, $V_{us},
V_{cb}, V_{ub}, m_u, m_d, m_s$ and $m_b$ are in good agreement with the
experiment when either {\it i) } $tan\beta \simeq 1$ or
{\it ii)} $tan\beta \geq 30$.  A correct prediction for
the bottom mass gives a lower limit on $m_t\geq 125$ for case {\it (i)}, in
agreement with the previous analysis, while in case {\it (ii)}  $m_t\geq
145$.
 \end{abstract}

\vspace*{6.0cm}

\begin{flushleft}
IOA 281/92 \\
NTUA 37/92 \\
December 1992
\end{flushleft}
\vfill
\newpage
%\vfill\eject
There is a firm belief among the high energy physicists that in the
ultimate theory, all the arbitrary parameters of the standard model
will be determined only from a small number of inputs at some
unification scale.

Recently, several attempts have been made $^{\cite{d,g,r,s}}$ to
determine the possible structures of the fermion mass matrices at the
Grand Unification(GUT) scale, which lead to the correct low energy mass
spectrum and to the maximal number of predictions. In Ref.\cite{g}, a
simple ansatz for the fermion mass matrices, at the GUT scale, was
proposed. The 13 arbitrary parameters of the low energy were
determined by 6 inputs, hence leading to 7 predictions.

In a recent paper $^{\cite{dlt}}$, the original ansatz of
Ref.\cite{g} for the fermion mass matrices  was expanded in order to
incorporate non-zero neutrino masses, as these are naturally predicted
in most of the GUT models. With
only 2  additional inputs at the GUT scale, it was found that all the previous
succesful predictions are retained, while one gets seven new low
energy predictions: the masses of the 3 light neutrinos, the leptonic
mixing angles and the corresponding CP-phase. It was also proposed
that the $m_{\nu _\tau}$ mass can give the correct contribution to the
hot dark matter, in agreement with the interpretation of the COBE data
$^{\cite{COBE}}$, while the solar neutrino problem
is solved via the MSW-mechanism~$^{\cite{neutrino}}$
 and $\nu _{\mu}\rightarrow \nu _e$
oscillations$^{\cite{solarn}}$.

All the previous calculations however, have been done for the case of
small bottom Yukawa coupling, compared to that of the top quark. This
corresponds to a relatively small ratio $\tan\beta$ of the two Higgs
vev's $<\bar h>$ and $<h>$ which give masses to the  up and down
quarks respectively. In this particular case one can ignore all but
the top Yukawa coupling corrections in the renormalization group
equations (RGEs) of the Yukawa couplings, and calculate to a good
approximation all the low energy parameters from the inputs at the GUT
scale.

In many unified models however - and in particular in string derived
GUTs - it is quite possible for the top and bottom quark Yukawa couplings
to be comparable at the GUT scale. In that case the bottom coupling
corrections play also an important r\^ole and should not be ignored.
Such a case would obviously correspond to a pretty large value of
$\tan \beta$, in contrast to the previous case.

In the present letter we are going to explore this latter case. We
start with an overview of the basic features of the proposed framework
$^{\cite{g}}$. It is assumed that there exists some Grand Unified
Supersymmetric Model (i.e. SO(10), SU(5), SU(4) etc) with the
following form of the mass matrices at the GUT scale
\begin{eqnarray}
M_u&=&\left(\begin{array}{ccc} 0&0&b\\0&b&0\\b&0&a\end{array} \right), \qquad
\quad
M_{\nu\nu^c}=\left(\begin{array}{ccc} 0&0&b\\0&b&0\\b&0&a\end{array} \right),
\label{eq:upmat}\\ &&  \nonumber \\
M_d&=& \left(\begin{array}{ccc} 0&f e^{i\phi}&0\\f e^{-i\phi}&d&2d\\0&2d&c
\end{array} \right), \quad
M_e = \left(\begin{array}{ccc} 0&f e^{i\phi}&0\\f e^{-i\phi}&-3d&2d\\0&2d&c
\end{array} \right), \label{eq:elecmat}\\ && \nonumber \\
&& \quad M_{\nu^c\nu^c}= M\, diag(k^{-2},k^{-1},1). \label{eq:majmat}
\end{eqnarray}
There is a factor of $-3$ difference in the $\{22\}$ entry of the matrix
$M_e$ compared to that of $M_d$. This arises naturally whenever these
entries are coupled to Higgs doublets belonging to specific
representations of the GUT group ($\underline{45}$ of SU(5) or
$\underline{126}$ of SO(10)). The relation of the $\{22\}$ and $\{23\}$
entries in  $M_d,M_e$  matrices is just a phenomenological assumption
$^{\cite{g}}$. The original ansatz for the $M_u$, $M_d$ and $ M_e$ matrices
was  augmented by a simple assumption for the Dirac, $M_{\nu\nu^c}$, and
the heavy Majorana, $M_{\nu^c\nu^c}$, neutrinos mass matrices:
$M_{\nu\nu^c}$ is simply taken to be identical to $M_u$, due to the
GUT relations, while $M_{\nu^c\nu^c}$ is taken for simplicity diagonal, whose
elements differ by a hierarchy factor $k\approx 10$. The predictions in the
neutrino sector have been discussed elsewhere${\cite{dlt}}$, thus we are not
going to elaborate them here.

The RGE's for the Yukawa couplings at the one loop level are
\begin{eqnarray}
16\pi^2 \frac{d}{dt} \lambda_U&=
& \left( I\cdot Tr [3 \lambda_U\lambda_U^\dagger ]  +
3 \lambda_U \lambda_U^\dagger +
\lambda_D \lambda_D^\dagger
-I\cdot G_U\right) \lambda_U, \label{eq:rge1}\\
16\pi^2 \frac{d}{dt} \lambda_N&=
& \left( I\cdot Tr [\lambda_U \lambda_U^\dagger ]  +
\lambda_E \lambda_E^\dagger
-I\cdot G_N\right) \lambda_N, \label{eq:rge2}\\
16\pi^2 \frac{d}{dt} \lambda_D&=
& \left( I\cdot Tr [3 \lambda_D\lambda_D^\dagger + \lambda_E\lambda_E^\dagger
]+
3 \lambda_D \lambda_D^\dagger +
\lambda_U \lambda_U^\dagger
-I \cdot G_D\right) \lambda_D, \label{eq:rge3}\\
16\pi^2 \frac{d}{dt} \lambda_E&=
& \left( I\cdot Tr [\lambda_E\lambda_E^\dagger +3 \lambda_D\lambda_D^\dagger ]+
3 \lambda_E \lambda_E^\dagger
-I \cdot G_E\right) \lambda_E,\label{eq:rge4}
\end{eqnarray}
where $\lambda_\alpha$, $\alpha=U,N,D,E$, represent the $3$x$3$ Yukawa matrices
which are defined in terms of the mass matrices given in
Eq.(\ref{eq:upmat}-\ref{eq:majmat}), and $I$ is the $3$x$3$ identity
matrix and
\begin{eqnarray}
G_\alpha&=&\sum_{i=1}^3 c_\alpha^i g_i^2(t),\\
g_i^2(t)&=&\frac{g_i^2(t_0)}{1- \frac{b_i}{8\pi^2}
g_i^2(t_0)(t-t_0)}\\
\{c_U^i \}_{i=1,2,3} &=& \left\{ \frac{13}{15},3,\frac{16}{3} \right\}, \qquad
\{c_D^i \}_{i=1,2,3} = \left\{\frac{7}{15},3,\frac{16}{3} \right\}, \\
\{c_E^i \}_{i=1,2,3} &=& \left\{ \frac{9}{5},3,0\right\}, \qquad \quad
\{c_N^i \}_{i=1,2,3} = \left\{ \frac{3}{5},3,0\right\}.
\end{eqnarray}
Following Ref.\cite{g} we diagonalize the up quark Yukawa matrix at
the GUT scale and redefine properly the lepton and quark fields
\begin{eqnarray}
\lambda_U\rightarrow {\tilde\lambda}_U&=& K^\dagger\lambda_UK,\qquad
{\lambda}_N
\rightarrow
{\tilde\lambda}_N= K^\dagger\lambda_NK, \nonumber \\
\lambda_D\rightarrow {\tilde\lambda}_D&=& K^\dagger\lambda_DK,\qquad
\lambda_E\rightarrow {\tilde\lambda}_E= K^\dagger\lambda_EK.  \label{eq:redef}
\end{eqnarray}
where now ${\tilde\lambda}_U$ is diagonal and $K$ is
\begin{equation}
K=\left( \begin{array}{ccc} \cos\theta & 0& \sin\theta \\ 0&1&0\\
-\sin\theta&0&\cos\theta \end{array} \right), \quad \tan 2\theta=\frac{2b}{a}.
\end{equation}
Now assuming that the only significant Yukawa terms are
${\tilde\lambda}_{U_{33}}$,  ${\tilde\lambda}_{D_{33}}$ and
${\tilde\lambda}_{E_{33}}$ , to a good approximation we may drop all other
terms in the parentheses of the  right-handed side of the RGE's above , and
write them formally as follows
 \begin{eqnarray}
{\tilde\lambda}_U(t)&=&\gamma _U
  \left(\begin{array}{ccc}1&0&0\\0&1&0\\0&0&{\zeta \xi^3}\end{array}\right)
               \xi^3{\tilde\lambda}_U(t_0)\label{eq:l-u}\\
{\tilde\lambda}_D(t)&=&\gamma _D
  \left(\begin{array}{ccc}1&0&0\\0&1&0\\0&0&\xi\zeta ^3\end{array}\right)
                \zeta^3 \zeta ^{\prime}{\tilde\lambda}_D(t_0)\label{eq:l-d}\\
{\tilde\lambda}_E(t)&=&\gamma _E
  \left(\begin{array}{ccc}1&0&0\\0&1&0\\0&0&\zeta ^{\prime 3}\end{array}\right)
                \zeta^3 \zeta ^{\prime}{\tilde\lambda}_E(t_0)\label{eq:l-e}\\
{\tilde\lambda}_N(t)&=&\gamma _N
  \left(\begin{array}{ccc}1&0&0\\0&1&0\\0&0&\zeta
^{\prime}\end{array}\right)
                \xi {\tilde\lambda}_N(t_0)\label{eq:l-n}
\end{eqnarray}
where
\begin{eqnarray}
\gamma_\alpha(t)&=& \exp({-\int G_\alpha(t) \,dt/(16\pi^2)})\\
%&=& \prod_{j=1}^3 \left( \frac{\alpha_{j,0}}{\alpha_j}\right)
%            ^{c_\alpha^j/2b_j}\\
%&=& \prod_{j=1}^3 \left(1- \frac{b_{j,0}\alpha_{j,0}(t-t_0)}{2\pi}
%\right)^{c_\alpha^j/2b_j},\\
\xi &=& \exp({\frac{1}{16\pi^2}\int_{t_0}^t {\tilde\lambda}_t dt}) \\
%&=&\left( 1-\frac{\kappa}{8\pi^2}\lambda_\alpha(t_0)
%\int_{t_0}^{t} \gamma_\alpha^2(t)\,dt \right)^{-1/(2\kappa)}.
\zeta &=& \exp({\frac{1}{16\pi^2}\int_{t_0}^t {\tilde\lambda}_b dt}) \\
\zeta^{\prime} &=& \exp({\frac{1}{16\pi^2}\int_{t_0}^t {\tilde\lambda}_\tau
dt})
\\
\end{eqnarray}
where ${\tilde\lambda}_t$, ${\tilde\lambda}_b$ and
${\tilde\lambda}_\tau$ stand for ${\tilde\lambda}_{U_{33}}$,
${\tilde\lambda}_{D_{33}}$ and ${\tilde\lambda}_{E_{33}}$ respectively.
In the above equations ${\tilde\lambda}_\tau$ is also included since
it satisfies the same initial condition with ${\tilde\lambda}_b$. Of
course the evolution down to $M_Z$ is different and $\zeta^\prime$
stays very close to 1 as long as the initial value of ${\tilde\lambda}_\tau$
 does not get too
large. Note that in the limit where
${\tilde\lambda}_t\gg {\tilde\lambda}_b$, ${\tilde\lambda}_\tau$ we
get $\zeta\approx \zeta^\prime\approx 1$ and the above equations reduce to the
simple uncoupled form
\begin{equation}
{\tilde\lambda}_\alpha (t)=\xi ^k\gamma _\alpha\lambda _\alpha (t_0)
\end{equation}
where now
\begin{equation}
\xi=\left(1-\frac{k}{8\pi ^2}{\tilde\lambda}_t(t_0)
   \int ^t_{t_0}\gamma^2_\alpha (t)dt\right)^{(\frac{-1}{2k})}
\end{equation}
and we recover the previous results $^{\cite{g,dlt}}$. In the general
case, however, the differential equations remain coupled and only a
numerical solution is possible.

We obtain the following relations among the masses
\begin{equation}
m_t=\zeta \xi ^3\frac{\eta _u m_c^2}{\eta^2_c m_u}\label{eq:mt}
\end{equation}
and for the down quarks and leptons
\begin{eqnarray}
m_b &\approx & \frac{\gamma _D}{\gamma _E}
\frac {\zeta^3}{\zeta^{\prime 3}}\xi m_\tau \eta _b
          \label{eq:mb}\\
m_s &\approx & \eta_s\frac{\gamma _D}{\gamma _E}\frac{m_\mu}{3}
      \left(1-\frac{4}{9}(1+3\zeta^{\prime
3})\frac{m_\mu}{m_\tau}\right) \label{eq:ms}\\
 m_d &\approx &
\eta_d\frac{\gamma_D}{\gamma_E}3m_e
      \left(1+\frac{4}{9}(1+3\zeta^{\prime
3})\frac{m_\mu}{m_\tau}\right)\label{eq:md}
\end{eqnarray}
where in the above relations $\eta_\alpha$, ($\alpha=b,c$) are taking
into account the QCD renormalization effects of the corresponding
quark masses from the energy scale $m_t$ down to their masses
$\eta_\alpha=m_\alpha(m_\alpha)/m_\alpha(m_t)$, while for
($\alpha=u,d,s$) we use
$\eta_\alpha = m_\alpha(1{\rm GeV})/m_\alpha(m_t)$. In what follows, we use the
values $\eta _b=1.4$, $\eta _c=1.8$ and $\eta _u=\eta _d=\eta _s=2.0$.
 The bottom quark mass is taken to lie in the range
$m_b(m_b)=4.25\pm.1$GeV. Thus, from the relation (26), the correct
prediction for $m_b$, fixes the combination  $\xi(\zeta /{\zeta ^{\prime }
})^3\approx .81\pm.02$ (the rest of the renormalization group parameters
involved in (26) vary slowly in terms of the input parameters). In order to
expess the predictions for the Kobayashi Maskawa (KM) mixing angles, we use the
following parametrization for the KM-matrix
\begin{eqnarray}
V_{KM}=\left( \begin{array}{ccc}
c_1c_3e^{i\phi}-s_1s_2s_3 & s_1c_3e^{i\phi}+c_1s_2s_3& -c_2s_3 \\
-s_1c_2 &c_1c_2& s_2 \\
c_1s_3e^{i\phi}+s_1s_2c_3&s_1s_3e^{i\phi}-c_1s_2c_3& c_2c_3 \end{array} \right)
\end{eqnarray}

where $c_1= \cos\theta_1$, $s_1=\sin\theta_1$, {\it etc}.
Then the predictions are:
\begin{eqnarray}
V_{us}&\simeq&3:s_1:\simeq\sqrt{\frac{m_e}{m_{\mu}}}
(1-\frac{25}{2} \frac{m_e}{m_{\mu}}+\frac {16}{9}\frac{m_{\mu}}{m_{\tau}}),
\label{eq:KMmix1}\\
V_{cb}&\simeq&:s_2:\simeq \frac{2}{3} (\xi \zeta ^3)^{-1}
 \frac{m_{\mu}}{m_{\tau}}(1-\frac{m_e}{m_{\mu}}-\frac{1}{9}
\frac{m_{\mu}}{m_{\tau}}) ,
\label{eq:KMmix2}\\
V_{ub}&\simeq&:s_3:\simeq (\frac{\xi}{\zeta})^2\frac{m_c}{\eta_c m_t} .
\label{eq:KMmix3}
\end{eqnarray}
{}In order to compute the various renormalization group parameters which
enter the various relations given above, we solve numerically the
equations (14-16) assuming the initial condition at $M_{GUT}\simeq
10^{16}GeV$, with $g_{GUT}\simeq \frac{1}{25.1}$. We are taking supersymmetric
beta function coefficients from $M_{GUT}$ down to $m_t$, while below $m_t$
we run the system with non-supersymmetric ones. We ensure that the gauge
couplings lie in the experimentally accepted region at $m_W$
and we compute the quark masses for a wide range of $tan\beta $, each time
using the proper initial values for the couplings $\lambda _{0,t},\lambda
_{0,b}$. Obviously, when $tan\beta \simeq 1$, a small ratio $r=\frac{\lambda
_{0,b}}{\lambda _{0,t}}$ is needed, while when $r\simeq 1$ then $tan\beta
\gg1$. Note however, that $\lambda _{0,b}=\lambda _{0,\tau}$, while $\lambda
_{0,\tau}$ and  $tan\beta $ should also be chosen so as to give the correct
$\tau $ mass. Thus, a consistency check is done for each chosen pair of
values ($\lambda _{0,\tau}, tan\beta $) seperately, where the $\tau $ mass is
taken to be $m_{\tau }=1784.1MeV$. Our numerical analysis reproduces the
previous results when $tan\beta \le 5$ and extends the analysis to the case
where $tan\beta \gg1$.

In order to see clearly the effect of a
large  $tan\beta $, (or equivallently a large $\lambda _{0,b}$ coupling), in
figure (1a) we plot the bottom mass versus the bottom coupling $\lambda
_{0,b}$ at the GUT scale, for constant top-mass $m_t=145GeV$, for three
successive approximations: Contour (I) represents the  case where  $\lambda
_{0,b},\lambda _{0,\tau}$-corrections are neglected. Contour (II) represents
the solution where only  $\lambda _{0,\tau}$ correction is neglected, while
case (III) is the contour which corresponds to the complete differential
system (14-16) where the corrections from all three couplings are taken
into account in the running. All curves are almost identical for small
$\lambda _b$ and $tan\beta <5$. Curve (I) is no longer valid for  $tan\beta
>5$ while curve (II) is not a good approximation for  $tan\beta >10$. Case
(III), but in terms of the ratio $r=\frac{\lambda _{0,b}}{\lambda _{0,t}}$
is shown in figure (1b). However, for this particular value of $m_t$,
 reasonably large values of $tan\beta (\sim 10)$, are excluded from the bottom
mass range which
%is restricted to lie in
is shown as the shaded region of these figures.
Acceptable $m_b$ values are possible only for $tan\beta \geq 40 (!)$, but
this corresponds to the unlikely case of $r>1$. Moreover, additional
constraints arise from the KM-mixing angles which are given in
(29-32). In particular, for the $m_t=145GeV$ case, the element $V_{bc}$
is in its upper experimentally allowed limit ($\simeq 5.4\times 10^{-2}$),
only when  $r\equiv \frac{\lambda _{0,b}}{\lambda _{0,t}}\simeq 1.1 $, while
-- unless $r\ll 1$ -- all the rest of the region of  $r$  is excluded either
from $m_b$ or from  $V_{bc}$ constraints. In
fact, the value $m_t\simeq 145GeV$, is the lower top mass which can be obtained
for a large $\lambda _b$ coupling, while smaller top mass values can be
obtained only for a neglegible  $\lambda _b$ coupling. In table I we collect
various results for the experimentally measured parameters for various values
of $m_t$. For all these cases we find that $m_d\approx 6.8MeV$ and $m_s\approx
138MeV$.

 As  $m_t$ gets higher
however, smaller $tan\beta $ values are possible. In figure (2) we show such a
case for $m_t=170GeV$ where we plot  $m_b$ versus $tan\beta $. A comparison of
the two curves in terms of the $\lambda _b$ values is presented in figure (3).
The shaded area (whose upper bound corresponds to $m_t\simeq (170\pm 3)GeV)$,
is prevented due to the bad ratio $\frac{m_u}{m_c}$. Thus, it is remarkable
that this ratio
 which is derived only in terms of the running $m_u, m_c$ masses determined
by well known methods$^{\cite{gl}}$, can put an upper bound on the top-quark
mass. If, on the other hand, from the electroweak breaking mechanism in
supesymmetric models we demand the condition $tan\beta>1$, then we
obtain a lower bound  $m_t\geq 125GeV$, although, this second bound is
reffered to a very small region as can be seen from figure (4).

In conclusion, we have reconsidered a
 proposed $^{\cite{g}}$ {\it ansatz } for the fermion mass
matrices at the GUT scale, and studied in detail the effects of a large bottom
Yukawa coupling on the various experimentally measured parameters of the low
energy theory. We have shown that the renormalization corrections of the
${\lambda _{b}}$ Yukawa coupling have a significant impact whenever
$r=\frac{\lambda _{0,b}}{\lambda _{0,t}}\geq 0.1$, while when $r\ll 1$,
they can safely be neglected. Furthermore, for  $r\geq 1$,
 $m_b$- and $V_{cb}$- low energy bounds
 put a lower limit on the top mass $m_t\geq 145GeV$, while in the case
of $r\ll 1$, one gets a less restrictive top mass $m_t\geq 125GeV$. In both
cases, an upper bound on the top mass can be obtained, $m_t \leq 173 GeV$,
from the relation (25).
% \vspace*{1.0cm}

\bigskip
The work of G.K.L. is partially supported by a C.E.C Science Program
SCI-0221-C(TT), while of N.D.T by C.E.C. Science Program
SC1-CT91-0729.

\vfill
%\eject
\newpage

\vspace*{.5cm}
TABLE I.

The predictions for the Ansatz of ref[2] for  large values of
$\lambda _{0,b}$ and $tan\beta $. The corresponding experimental ranges are
$V_{ub}=(4\pm3)\times 10^{-3}$,$V_{cb}=(4.3\pm1.1)\times
10^{-2}$.$^{\cite{PD}}$ From
(27-30) we also get  $m_d\approx 6.8MeV$ ,  $m_s\approx 138MeV$ and
$V_{ub}\approx .218$, in agreement with the experiment.

\vspace*{1cm}
\begin{tabular}{||c|l|c|c|c|c|c||}    \hline \hline
$m_t(GeV)$   &\ \ \ \ ${}r$   &$V_{ub}\times 10^{3}$  &$V_{cb}\times
10^{2}$  &$m_b(GeV)$ &$tan\beta $  & {}    \\  \hline \hline
 145     &$\sim 1.1$  &$4.90\pm 0.20$ &$5.40$&4.35& 45 &$\surd$\\
\cline{2-7}
         &$\sim 1.3$&$5.12\pm 0.20$ &$5.91$&4.26& 50  &{}\\   \hline
155     &$\sim 0.83$  &$4.30\pm 0.17$ &$5.40$&4.31& 45  &$\surd$\\
\cline{2-7}
         &$\sim 1.0$&$4.53\pm 0.18$ &$5.98$&4.20 & 50 &{}\\   \hline
165     &$\sim 0.35$  &$3.58\pm 0.14$ &$5.02$&4.34& 35 &$\surd$\\
\cline{2-7}
         &$\sim 0.63$&$3.75\pm 0.15$ &$5.56$&4.19 & 45 &{}\\   \hline
170     &$\sim 0.33$  &$3.25\pm 0.15$ &$4.91$&4.32& 30  &$\surd$\\
\cline{2-7}
         &$\sim 0.38$&$4.39\pm 0.17$ &$5.57$&4.26& 35  &{}\\ \hline \hline
\end{tabular}

\vspace*{5cm}

{\bf Figure Captions}
\vspace*{1cm}

{\bf Fig.1}. Bottom mass for $m_t=145GeV$, {\it a)} as a function of $\lambda
_{0,b}(M_{GUT})$, and {\it b)} as a function of
$r=\frac{\lambda _{0,b}}{\lambda _{0,t}}$(see text for details).

{\bf Fig.2}. Bottom mass for $m_t=170GeV$ as a function of $\tan\beta $.

{\bf Fig.3}. Bottom mass for $m_t=145$ and $170 GeV$. The shaded area is
prevented due to unacceptable $\frac {m_u}{m_c}$ ratio.

\vfill\eject

\end{document}